\newcounter{myctr}
\def\myitem{\refstepcounter{myctr}\bibfont\noindent\ifnum\themyctr>9\else\phantom{0}\fi\hangindent17pt\themyctr.\enskip}

\documentclass{ws-ijqi}
\usepackage{graphicx,color}
\usepackage{amsmath}
\usepackage{amssymb}
\usepackage{bm}

\begin{document}

\markboth{S. Ando, K. Yuasa and M. Iazzi}
{Interference of an Array of Independent Bose-Einstein Condensates}

\catchline{}{}{}{}{}

\title{INTERFERENCE OF AN ARRAY OF INDEPENDENT BOSE-EINSTEIN CONDENSATES WITH FIXED NUMBER OF ATOMS}

\author{SATOSHI ANDO}
\address{Department of Physics, Waseda University, Tokyo 169-8555, Japan}

\author{KAZUYA YUASA}
\address{Waseda Institute for Advanced Study, Waseda University, Tokyo 169-8050, Japan\\
yuasa@aoni.waseda.jp}

\author{MAURO IAZZI}
\address{International School for Advanced Studies (SISSA), via Beirut 2-4, I-34014 Trieste, Italy\\
mauro.iazzi@sissa.it}

\maketitle

\begin{history}
\received{28 May 2010}
\revised{2 April 2011}
\end{history}

\begin{abstract}
Interference of an array of independent Bose-Einstein condensates, whose experiment has been performed recently, is theoretically studied in detail.
Even if the number of the atoms in each gas is kept finite and the phases of the gases are not well defined, interference fringes are observed on each snapshot.
The statistics of the snapshot interference patterns, i.e., the average fringe amplitudes and their fluctuations (covariance), are computed analytically, and concise formulas for their asymptotic values for long time of flight are derived.
Processes contributing to these quantities are clarified and the relationship with the description on the basis of the symmetry-breaking scenario is revealed.
\end{abstract}

\keywords{Bose-Einstein condensation; interference; Hanbury Brown and Twiss effect; snapshot; spontaneous symmetry breaking.}

\section{Introduction}
Interference fringes are observed when two independently prepared Bose-Einstein condensates (BECs) are released and overlap after a time of flight.\cite{ref:InterferenceBEC-IJQI}
The simplest explanation of this phenomenon is based on the spontaneous symmetry breaking of BEC, where it is assumed that the phases of the two Bose gases are individually fixed upon condensation, which enable the gases to interfere with each other.\cite{ref:BECPitaevskiiStringari-IJQI,ref:BEC-reviews}

When the number of the bosonic atoms in each gas is finite, however, the symmetry of the system can never be broken.
Javanainen and Yoo pointed out in their seminal paper\cite{ref:JavanainenYoo-IJQI} that, even if the number of the atoms in each gas is well defined and its phase is accordingly uncertain, the two gases can exhibit interference.
Note that the quantum mechanics predicts result of accumulation of many experiments.
We usually look at the single-particle distribution to study interference in the density profile, which predicts the image obtained by accumulating many photos of the overlapping BECs.
No interference can be observed in this quantity if the number of the atoms in each gas is fixed and its phase is not well defined.
However, Javanainen and Yoo demonstrated that interference fringes show up on \textit{each snapshot}.\cite{ref:JavanainenYoo-IJQI}
It is important to notice that we look at many identical atoms at the same time by taking a photo of the cloud.
The symmetrization of the wave function gives rise to correlations among the multiple atoms in the cloud (Hanbury Brown and Twiss effect\cite{ref:Loudon-IJQI}), which make the probability to find an atom in the presence of other identical atoms non-unform in space.
This is the origin of the oscillation in the density profile on each snapshot.
Such an interference is called ``measurement-induced interference''\cite{ref:BEC-reviews,ref:JavanainenYoo-IJQI,ref:MeasurementInduced,ref:CastinDalibard-IJQI,ref:BECPethickSmith-IJQI} since measurement (taking a photo) extracts an interference pattern.

An experiment was carried out to study the interference of an \textit{array} of independent BECs.\cite{ref:Hadzibabic-BECIntArray-IJQI}
Note that the snapshot interference patterns differ from shot to shot.\footnote{In particular, the fringe pattern shifts from run to run, and the fringes are smeared out if all the snapshots are superimposed, reproducing the quantum-mechanical expectation for single-particle distribution.}
A certain number of snapshots were collected by repeating the experiment, and the fluctuations of the fringe amplitudes were analyzed, on the basis of the hypothesis of the broken symmetry of the BECs.
The interference of the array of BECs is briefly discussed in Ref.\ \refcite{ref:PolkovnikovEPL-IJQI}, concentrating on the fluctuation of a measure of interference, with each BEC consisting of a fixed number of atoms, without assuming the symmetry breaking.
In this paper, we set up tools for studying the snapshot interference.
These are essentially equivalent to the ones suggested in Refs.\ \refcite{ref:PolkovnikovEPL-IJQI} and \refcite{ref:Boston}, where the interference of the array of BECs, each containing a fixed number of atoms, is studied.
The time-evolution of the fringe amplitudes expected on snapshots and their fluctuations (covariance) are computed, and concise formulas for the average fringe amplitudes and their covariance are derived, which characterize the statistics of the snapshot interference patterns after long time of flight.
The processes contributing to these quantities are clarified and the relationship with the description on the basis of the symmetry-breaking scenario, which well explains the experimental data, is revealed.

\section{Snapshot Profile and Its Fluctuation}
Suppose that there are $N$ identical bosonic atoms and one takes a ``photo'' of the cloud.
The probability of finding the $N$ atoms at positions $\{\bm{r}_1,\ldots,\bm{r}_N\}$ at an instant $t$ is given by the $N$-particle probability distribution function
\begin{equation}
P_t^{(N)}(\bm{r}_1,\ldots,\bm{r}_N)
=\frac{1}{N!}\langle\hat{\psi}^\dag(\bm{r}_1)\cdots\hat{\psi}^\dag(\bm{r}_N)\hat{\psi}(\bm{r}_N)\cdots\hat{\psi}(\bm{r}_1)\rangle_t,
\label{eqn:PN}
\end{equation}
where $\hat{\psi}(\bm{r})$ is the field operator of the bosonic atom, satisfying the canonical commutation relations
\begin{equation}
[\hat{\psi}(\bm{r}),\hat{\psi}^\dag(\bm{r}')]=\delta^3(\bm{r}-\bm{r}'),\quad\text{etc.},
\end{equation}
and the average $\langle\cdots\rangle_t$ is taken over the state of the cloud evolved from time $0$ to $t$, while the operators are time-independent (Schr\"odinger picture).
This probability is normalized to unity,
\begin{equation}
\int d^3\bm{r}_1\cdots d^3\bm{r}_N\,P_t^{(N)}(\bm{r}_1,\ldots,\bm{r}_N)=1.
\end{equation}
Note the relation 
\begin{equation}
\int d^3\bm{r}_\ell\,P_t^{(N)}(\bm{r}_1,\ldots,\bm{r}_\ell,\ldots\bm{r}_N)
=P_t^{(N-1)}(\bm{r}_1,\ldots,\bm{r}_{\ell-1},\bm{r}_{\ell+1},\ldots\bm{r}_N).
\end{equation}
Given a single configuration $\{\bm{r}_1,\ldots,\bm{r}_N\}$, the \textit{snapshot} density profile of the cloud is constructed as
\begin{equation}
\rho(\bm{r})=\frac{1}{N}\sum_{i=1}^N\delta(\bm{r}-\bm{r}_i).
\end{equation}
This profile is also normalized as
\begin{equation}
\int d^3\bm{r}\,\rho(\bm{r})=1.
\label{eqn:DensityNormalization}
\end{equation}

Notice that the specific configuration $\{\bm{r}_1,\ldots,\bm{r}_N\}$ appears according to the probability $P_t^{(N)}(\bm{r}_1,\ldots,\bm{r}_N)$ given in (\ref{eqn:PN}) and the snapshot profile $\rho(\bm{r})$ fluctuates from run to run.
The average profile
\begin{equation}
\overline{\rho(\bm{r})}
=\int d^3\bm{r}_1\cdots d^3\bm{r}_N\,P_t^{(N)}(\bm{r}_1,\ldots,\bm{r}_N)\rho(\bm{r})
=P_t^{(1)}(\bm{r}),
\label{eqn:P1}
\end{equation}
which represents the image obtained by accumulating many snapshots, coincides with the single-particle probability distribution $P_t^{(1)}(\bm{r})$ [We omit to specify the time dependence of the averages $\overline{\vphantom{|}\cdots\vphantom{|}}$, which depend on the time specified in $P_t^{(N)}$; $\rho(\bm{r})$ itself is time-independent].
However, $\overline{\rho(\bm{r})}$ is not the quantity of interest.
We are interested in the presence of interference fringes on \textit{each} snapshot $\rho(\bm{r})$, while the single-particle distribution $P_t^{(1)}(\bm{r})$ (superposition of many snapshots) does not exhibit interference pattern.\cite{ref:JavanainenYoo-IJQI}
In order to detect an oscillating pattern on each snapshot, we look at the Fourier spectrum of the density profile $\rho(\bm{r})$,
\begin{equation}
\tilde{\rho}(\bm{k})
=\int d^3\bm{r}\,\rho(\bm{r})e^{-i\bm{k}\cdot\bm{r}}.
\end{equation}
Possible nontrivial spikes in $\tilde{\rho}(\bm{k})$, representing the presence of interference fringes \textit{in each snapshot} would disappear if average is taken over all the snapshots.
The reason for the disappearance of the interference fringes is the random shift of the interference pattern from snapshot to snapshot, which smears out the fringes.\cite{ref:JavanainenYoo-IJQI}
We can discard such random offset by removing the phase of the Fourier spectrum, i.e., by looking at the (square) modulus $|\tilde{\rho}(\bm{k})|^2$.
The fringe spikes would then survive on average in the quantity
\begin{equation}
S_t(\bm{k})
=\overline{|\tilde{\rho}(\bm{k})|^2},
\label{eqn:Sdef}
\end{equation}
if the majority of the snapshots exhibit interference patterns with a definite fringe spacing, irrespectively of the random spatial shifts.
The fluctuation of the fringe spectrum from snapshot to snapshot is estimated by the variance, or more generally, by the covariance
\begin{equation}
C_t(\bm{k},\bm{k}')
=\overline{|\tilde{\rho}(\bm{k})|^2|\tilde{\rho}(\bm{k}')|^2}-\overline{|\tilde{\rho}(\bm{k})|^2}\cdot\overline{|\tilde{\rho}(\bm{k}')|^2}.
\label{eqn:Cdef}
\end{equation}
These are our tools for studying the snapshot interference patterns.
Notice that we could define an observable representing a (relative) phase $\theta(\bm{k}) = \arg\tilde{\rho}(\bm{k})$, which is \textit{not} conjugated to the (relative) density.

By noting
\begin{equation}
\tilde{\rho}(\bm{k})
=\frac{1}{N}\sum_{i=1}^Ne^{-i\bm{k}\cdot\bm{r}_i},
\end{equation}
one realizes that these quantities are controlled by few-particle distribution functions.
Indeed, one gets
\begin{gather}
S_t(\bm{k})=\frac{N-1}{N}I_t^{(2)}(\bm{k})+\frac{1}{N},
\label{eqn:SI}
\\
C_t(\bm{k},\bm{k}')=\frac{(N-1)!}{N^3(N-4)!}I_t^{(4)}(\bm{k},\bm{k}')
-\frac{(N-1)^2}{N^2}I_t^{(2)}(\bm{k})I_t^{(2)}(\bm{k}')+O\!\left(\tfrac{1}{N}\right),
\label{eqn:CI}
\end{gather}
where
\begin{gather}
I_t^{(2)}(\bm{k})
=\int d^3\bm{r}_1\, d^3\bm{r}_2\,
P_t^{(2)}(\bm{r}_1,\bm{r}_2)
 e^{ i\bm{k}\cdot(\bm{r}_1-\bm{r}_2)},
\label{eqn:I2def}
\\
I_t^{(4)}(\bm{k},\bm{k}')
=
\int
d^3\bm{r}_1\, d^3\bm{r}_2\,d^3\bm{r}_3\, d^3\bm{r}_4\,
P_t^{(4)}(\bm{r}_1,\bm{r}_2,\bm{r}_3,\bm{r}_4)
 e^{ i\bm{k}\cdot(\bm{r}_1-\bm{r}_2)+ i\bm{k}'\cdot(\bm{r}_3-\bm{r}_4)}.
 \label{eqn:I4def}
\end{gather}
The interference pattern of the $N$ particles is essentially ruled by the two-particle distribution $P_t^{(2)}(\bm{r}_1,\bm{r}_2)$, while its fluctuation by $P_t^{(4)}(\bm{r}_1,\bm{r}_2,\bm{r}_3,\bm{r}_4)$.
The formulas (\ref{eqn:SI}) and (\ref{eqn:CI}) show that the present formulation is essentially equivalent to the one employed in Ref.\ \refcite{ref:Boston}, when the number $N$ of the atoms in the cloud is large.

Note the symmetries of these quantities, $S_t(\bm{k})=S_t(-\bm{k})$, $C_t(\bm{k},\bm{k}')=C_t(\bm{k}',\bm{k})=C_t(\bm{k},-\bm{k}')=C_t(-\bm{k},\bm{k}')=C_t(-\bm{k},-\bm{k}')$.

\section{Interference of an Array of Bose-Einstein Condensates}
Let us now study the interference of an array of independent BECs.
Suppose that $K$ independent BECs are created at regular distances in a periodic lattice with a lattice constant $d$, forming a 1D array of BECs along the $x$ axis (in the actual experiment, the array is loosely confined in the transverse directions and the BECs are like disks).
We assume that each BEC contains exactly $N$ atoms of mass $m$, which are all condensed in the ground state of each well at zero temperature.
We consider ideal gases, neglecting the intra-atomic interactions.
The wells are well separated with vanishing tunneling probability between adjacent wells.
We emphasize that these BECs are mutually independent, with the fixed number of atoms in each BEC endowed with no definite (relative) phase, and there is no phase correlation among them.
At time $t=0$, the optical lattice is switched off and the array of BECs is released.
After free expansion in space (we do not consider gravitational field), a photo of the cloud is taken to observe an interference pattern among the BECs.
The expansions of each BEC in the longitudinal and transverse directions are decoupled, and in addition, the expansion in the transverse directions is slower than that in the longitudinal direction, since the transverse confinement is weaker than in the longitudinal direction.
In the following analysis, we concentrate on the dynamics in the longitudinal direction, which is relevant to the appearance of the interference fringes.

Let $\varphi_n(x)$ ($n=1,\ldots,K$) denote the wave function of the ground state of the $n$th potential well and $\hat{a}_n$ the associated annihilation operator.
We assume that the wave functions are orthogonal to each other,
\begin{equation}
\int dx\,\varphi_n^*(x)\varphi_{n'}(x)=\delta_{nn'},
\label{eqn:Orthogonality}
\end{equation}
and the bosonic operators of different wells commute,
\begin{equation}
[\hat{a}_n,\hat{a}_{n'}^\dag]=\delta_{nn'},\quad\text{etc.}
\end{equation}
During the time of flight, the wave functions evolve in free space according to
\begin{equation}
\varphi_n(x,t)=e^{\frac{i\hbar t}{2m}\frac{\partial^2}{\partial x^2}}\varphi_n(x).
\end{equation}

Let us first look at the average of the snapshot profiles, $\overline{\rho(x)}$, which gives the single-particle distribution of the atoms in the cloud, $P_t^{(1)}(x)$, as shown in (\ref{eqn:P1}).
It is given by
\begin{equation}
\overline{\rho(x)}=P_t^{(1)}(x)=\frac{1}{K}\sum_{n=1}^K|\varphi_n(x,t)|^2,
\end{equation}
in which no interference fringes are observed.
This is the definition of the ``independence'' of the BECs.
Interference patterns are however found on each snapshot.
Let us look at the average spectrum $S_t(k)$ and the fluctuation $C_t(k,k')$ of the snapshot profiles, introduced in (\ref{eqn:Sdef})--(\ref{eqn:Cdef}) and evaluated by (\ref{eqn:SI})--(\ref{eqn:CI}).
Since the two-particle distribution of the atoms after the time of flight is given by
\begin{align}
P_t^{(2)}(x_1,x_2)
=
\frac{KN}{KN-1}\,\Biggl(&
P_t^{(1)}(x_1)
P_t^{(1)}(x_2)
-\frac{1}{K^2N}
\sum_{n=1}^K
|\varphi_n(x_1,t)|^2|\varphi_n(x_2,t)|^2
\nonumber\\
&{}+\frac{1}{K^2}\mathop{\sum_{n_1=1}^K\sum_{n_2=1}^K}_{n_1\neq n_2}
\varphi_{n_2}^*(x_1,t)\varphi_{n_1}^*(x_2,t)
\varphi_{n_2}(x_2,t)\varphi_{n_1}(x_1,t)
\Biggr),
\end{align}
its Fourier transformation (\ref{eqn:I2def}) yields
\begin{equation}
I_t^{(2)}(k)
=
\frac{KN}{KN-1}\,\Biggl(
|I_t^{(1)}(k)|^2-\frac{1}{K^2N}\sum_{n=1}^K
|\chi_{nn}(k,t)|^2
+
\frac{1}{K^2}\mathop{\sum_{n_1=1}^K\sum_{n_2=1}^K}_{n_1\neq n_2}
|\chi_{n_1n_2}(k,t)|^2
\Biggr),
\label{eqn:I2}
\end{equation}
where 
\begin{equation}
I_t^{(1)}(k)=\frac{1}{K^2}\sum_{n=1}^K\chi_{nn}(k,t),\quad
\chi_{n_1n_2}(k,t)
=\int dx\,\varphi_{n_1}^*(x,t)e^{-ikx}\varphi_{n_2}(x,t)
\label{eqn:chi}
\end{equation}
are the Fourier transforms of the single-particle distribution $P_t^{(1)}(t)$ and of the so-called ``interference term,'' respectively.
$I_t^{(4)}(k,k')$ defined in (\ref{eqn:I4def}) is also composed of $\chi_{ij}(k,t)$ as
\begin{align}
I_t^{(4)}(k,k')
={}&\frac{(KN-4)!}{(KN)!}
\nonumber
\displaybreak[0]\\
&{}\times\!
\sum_{n_1,n_2,n_3,n_4}
\sum_{n_1',n_2',n_3',n_4'}
\prod_i\delta_{(\delta_{n_1i}+\delta_{n_2i}+\delta_{n_3i}+\delta_{n_4i})(\delta_{n_1'i}+\delta_{n_2'i}+\delta_{n_3'i}+\delta_{n_4'i})}
\nonumber\displaybreak[0]\\
&\quad
{}\times 
\sqrt{N(N-\delta_{n_1n_2})(N-\delta_{n_1n_3}-\delta_{n_2n_3})(N-\delta_{n_1n_4}-\delta_{n_2n_4}-\delta_{n_3n_4})}
\nonumber\\
&\quad
{}\times 
\sqrt{N(N-\delta_{n_1'n_2'})(N-\delta_{n_1'n_3'}-\delta_{n_2'n_3'})(N-\delta_{n_1'n_4'}-\delta_{n_2'n_4'}-\delta_{n_3'n_4'})}
\nonumber\\
&\quad
{}\times 
\chi_{n_1n_1'}^*(k,t)\chi_{n_2'n_2}(k,t)
\chi_{n_3n_3'}^*(k',t)\chi_{n_4'n_4}(k',t).
\label{eqn:I4}
\end{align}

\subsection{Gaussian Wave Packets}
Let us look explicitly at the time-evolution of the interference of the array of BECs.
In the experiment reported in Ref.\ \refcite{ref:Hadzibabic-BECIntArray-IJQI}, $K=30$ condensates of $^{87}\text{Rb}$, each containing $N\sim10^4$ atoms, were created with a lattice constant $d=2.7\,\mu\text{m}$.
Each condensate can be regarded as being confined in a harmonic potential with frequency $\omega/2\pi=4\,\text{kHz}$ in the longitudinal direction (while with $\omega_\perp/2\pi=74\,\text{Hz}$ in the transverse directions), and its wave function in the longitudinal direction $x$ would be well approximated by a Gaussian 
\begin{equation}
\varphi_n(x)=\frac{1}{\sqrt[4]{2\pi(\Delta x)^2}}e^{-(x-nd)^2/4(\Delta x)^2}\qquad
(n=1,\ldots,K)
\end{equation}
of size $\Delta x=\sqrt{\hbar/2m\omega}=120\,\text{nm}$.
The overlap between the Gaussian wave functions of different wells,
\begin{equation}
\int dx\,\varphi_{n_1}^*(x)\varphi_{n_2}(x)=e^{-(n_1-n_2)^2d^2/8(\Delta x)^2},
\label{eqn:NonOrthogonality}
\end{equation}
is negligibly small since $\Delta x\ll d$, and they are approximately orthogonal to each other [see (\ref{eqn:Orthogonality})].

For these Gaussian wave functions, the Fourier transform of the interference term $\chi_{n_1n_2}(k,t)$ defined in (\ref{eqn:chi}) is given by
\begin{equation}
\chi_{n_1n_2}(k,t)
=e^{-(\Delta x)^2k^2/2}e^{-\omega^2t^2(\Delta x)^2[k-(n_1-n_2)md/\hbar t]^2/2}
e^{-i(n_1+n_2)kd/2},
\label{eqn:ChiGauss}
\end{equation}
and by inserting it to $I_t^{(2)}(k)$ in (\ref{eqn:I2}), we get the average spectrum $S_t(k)$ at an instant $t$,
\begin{align}
S_t(k)
={}
e^{-(\Delta x)^2k^2}&\biggl[
e^{-\omega^2t^2(\Delta x)^2k^2}
\left(
\frac{\sin^2(Kkd/2)}{K^2\sin^2(kd/2)}
-\frac{1}{KN}
\right)
\nonumber\\
&
{}+
\sum_{n=1}^{K-1}\frac{K-n}{K^2}
\Bigl(
e^{-\omega^2 t^2(\Delta x)^2(k-k_n)^2}
+e^{-\omega^2t^2(\Delta x)^2(k+k_n)^2}
\Bigr)
\biggr]
+\frac{1}{KN},
\label{eqn:SGauss}
\end{align}
where 
\begin{equation}
k_n=n\frac{md}{\hbar t}.
\label{eqn:kn}
\end{equation}
This expression is exact and valid for any $N$ and $t$.

\begin{figure}[b]
\begin{center}
\includegraphics{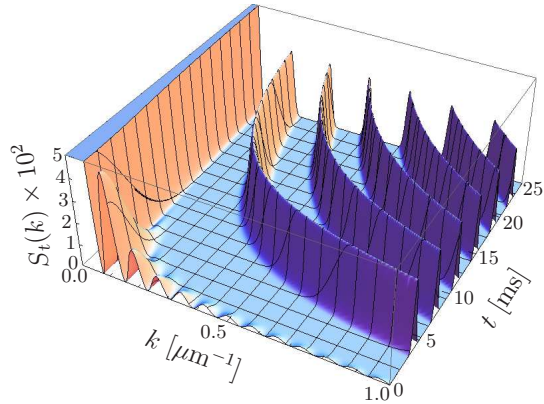}
\end{center}
\caption{The time-evolution of the average spectrum $S_t(k)$ of the interference of an array of Gaussian BECs, with the experimental parameters $K=30$, $N=10^4$, $d=2.7\,\mu\text{m}$, $\Delta x=120\,\text{nm}$, $\omega/2\pi=4\,\text{kHz}$.}
\label{fig:S}
\end{figure}

The time-evolution of the average spectrum $S_t(k)$ in (\ref{eqn:SGauss}) is shown in Fig.\ \ref{fig:S} with the experimental parameters recapitulated above.
\begin{figure}[t]
\begin{center}
\includegraphics{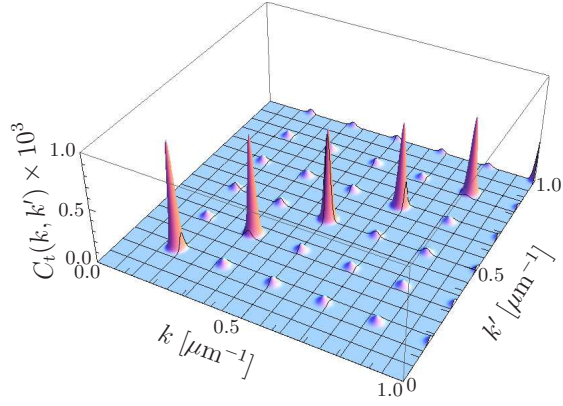}
\end{center}
\caption{The covariance $C_t(k,k')$ of the spectrum of the interference of an array of Gaussian BECs at time $t=22\,\text{ms}$, with the same experimental parameters as in Fig.\ \ref{fig:S}.}
\label{fig:C}
\end{figure}
After a short transient period, the  average spectrum $S_t(k)$ exhibits $2K-1=59$ peaks.
The central peak at $k=0$ represents the overall profile of the whole cloud, which becomes narrower and narrower as the cloud expands during the time of flight, while its height is kept constant $S_t(0)=1$, since the density profile $\rho(x)$ is normalized to unity in (\ref{eqn:DensityNormalization}).\footnote{$S_t(k)$ given in (\ref{eqn:SGauss}) for the Gaussian wave packets yields $S_t(0)\neq1$, contradicting with the normalization condition $S_t(0)=1$ in (\ref{eqn:DensityNormalization}). This is due to the nonorthogonality of the Gaussian wave functions $\varphi_n(x)$ in (\ref{eqn:NonOrthogonality}). The deviation of $S_t(0)$ from unity is actually of order of the overlaps among the Gaussian wave functions and can be neglected in the regime $\Delta x\ll d$ assumed in the present analysis.}
The side peaks at $k\simeq \pm k_n$ ($n=1,\ldots,K-1$), on the other hand, represent the interference fringes in the snapshot profiles.
Those peaks shift toward the center, according to the formula for $k_n$ in (\ref{eqn:kn}), while they become sharper and higher as time goes on.
In the experiment reported in Ref.\ \refcite{ref:Hadzibabic-BECIntArray-IJQI}, snapshots of the density profiles were taken at $t=22\,\text{ms}$ after the release from the optical lattice.
The covariance $C_t(k,k')$ of the interference spectrum at this time is shown in Fig.\ \ref{fig:C}, for a large $N$.

\subsection{Asymptotic Behavior}
The asymptotic heights of the peaks in $S_t(k)$ (Fig.\ \ref{fig:S}) and those in $C_t(k,k)$ (Fig.\ \ref{fig:C}) for large $t$ can be estimated without assuming the Gaussian wave packets.
Observe the asymptotic behavior of the wave packet for large $t$,
\begin{align}
\varphi_n(x,t)
&=\int dk\,e^{-i\hbar k^2t/2m}e^{ik(x-nd)}\tilde{\varphi}(k)\nonumber\\
&\sim\sqrt{\frac{m}{i\hbar t}}e^{im(x-nd)^2/2\hbar t}
\tilde{\varphi}\!\left(\frac{m}{\hbar t}(x-nd)\right).
\end{align}
Then, the Fourier transform of the interference term defined in (\ref{eqn:chi}),
\begin{align}
\chi_{n_1n_2}(k,t)
\sim{}&e^{-i(n_1^2-n_2^2)md^2/2\hbar t}
\nonumber\\
&{}\times\int dk'\,\tilde{\varphi}^*\!\left(k'-n_1\frac{md}{\hbar t}\right)\tilde{\varphi}\!\left(k'-n_2\frac{md}{\hbar t}\right)e^{-i(\hbar t/m)[k-(n_1-n_2)md/\hbar t]k'},
\label{eqn:AsympChi}
\end{align}
becomes sharply peaked at $k=(n_1-n_2)md/\hbar t$ with unit height $\chi_{n_1n_2}(k,t)
\sim e^{-i(n_1^2-n_2^2)md^2/2\hbar t}$, and $I_t^{(2)}(k)$ given in (\ref{eqn:I2}) develops $2K-1$ peaks
\begin{equation}
I_t^{(2)}(k)\sim\begin{cases}
\medskip
\displaystyle
1&\text{at}\quad k=0,\\
\displaystyle
\frac{K-n}{K^2(1-1/KN)}&\text{at}\quad k=\pm k_n\ (n=1,\ldots,K-1),
\end{cases}
\end{equation}
where $k_n$ is defined in (\ref{eqn:kn}).
At these wave numbers, $I_t^{(4)}(k,k')$ given in (\ref{eqn:I4}) is estimated to be
\begin{equation}
I_t^{(4)}(k_m,k_n)
\sim\begin{cases}
\medskip
\displaystyle
I_t^{(2)}(k_n)
\hfill
(m=0,n\ge0),\\
\displaystyle
\frac{N^4(KN-4)!}{(KN)!}
\left[
(K-n)\left(
2K-2n-1
-\frac{6}{N}
+\frac{1}{N^2}
\right)
+\frac{4n}{N}
\right]
\\
\medskip
\hfill
(m=n>0),\\
\displaystyle
\frac{N^4(KN-4)!}{(KN)!}
\left[
(K-m)(K-n)
+2(K-m-n)\left(
1-\frac{1}{N}
\right)
-\frac{n}{N}
\right]\\
\hfill
(m>n>0),
\end{cases}
\end{equation}
for large $t$.
Therefore, combining these expressions, we get concise expressions for the peaks in the average spectrum $S_t(k)$ and their fluctuations $C_t(k,k')$ for large $t$,
\begin{equation}
S_t(k_n)\sim
\begin{cases}
\medskip
1&(n=0),\\
\displaystyle
\frac{K-n}{K^2}+\frac{1}{KN}&(n>0),
\end{cases}
\label{eqn:PeakFormulaS}
\end{equation}
\begin{equation}
C_t(k_m,k_n)\sim
\begin{cases}
\medskip
0&(m=0,n\ge0),\\\medskip
\displaystyle
\frac{(K-n)(K-n-1)}{K^4}+O\!\left(\frac{1}{N}\right)&(m=n>0),\\
\displaystyle
\frac{2(K-m-n)}{K^4}+O\!\left(\frac{1}{N}\right)&(m>n>0).
\end{cases}
\label{eqn:PeakFormulaC}
\end{equation}
Remember the symmetry $C_t(k,k')=C_t(k',k)$.

The peaks of $S_t(k)$ at $k\neq0$ stem from the ``interference terms'' $\chi_{n_1n_2}(k,t)$ in the last term of $I_t^{(2)}(k)$ in (\ref{eqn:I2}).
This represents the interference of the atoms originating from the $n_1$th and $n_2$th condensates, and it exhibits a peak at $k=(n_1-n_2)md/\hbar t$, as shown in (\ref{eqn:AsympChi}) [see also (\ref{eqn:ChiGauss})].
This is because the atoms interfering between these condensates at time $t$ are those which have travelled the distance $|n_1-n_2|d/2$ in time $t$, i.e., they have velocities $\mp(n_1-n_2)d/2t$, and accordingly momenta $\mp(n_1-n_2)md/2t$.
The interference between the atoms counter-propagating with these momenta exhibits oscillation with wave number $k=(n_1-n_2)md/\hbar t$.
There are $K-1$ different distances between condensates, and hence there are $2(K-1)$ nontrivial peaks representing the interference fringes\footnote{Note that the density profile $\rho(x)$ is a real function and its Fourier spectrum $|\tilde{\rho}(k)|^2$ is an even function of $k$.} plus the peak at $k=0$.
Therefore, there are $2K-1$ peaks in the spectrum $S_t(k)$.

The peak at $k=k_n$ represents the interference of particles originating from pairs of condensates which are separated by a distance $nd$.
There are $K-n$ such pairs of sources contributing this interference, among which there is no phase correlation.
That is why the spectrum $S_t(k_n)$ is proportional to $K-n$, without interference among the contributions from different pairs of condensates, while its denominator $K^2$ is due to the normalization of the density profile.
On the other hand, the fluctuation $C_t(k,k')$ is controlled by the number of independent ways of selecting two pairs of condensates separated by a distance $nd$, which is given by $(K-n)(K-n-1)$.

\begin{figure}[t]
\begin{center}
\includegraphics[width=0.55\textwidth]{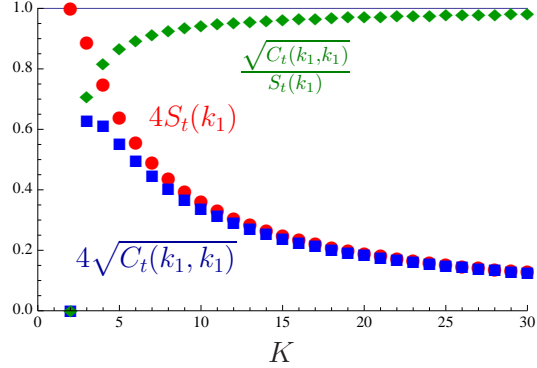}
\end{center}
\caption{The asymptotic strength of the average spectrum $S_t(k)$ at $k=k_1$ and its fluctuation $C_t(k_1,k_1)$, as functions of the number of condensates $K$, in the large $N$ limit.}
\label{fig:Peaks}
\end{figure}
In Fig.\ \ref{fig:Peaks}, the asymptotic strength of the average spectrum $S_t(k)$ at $k=k_1$ (interference between adjacent condensates) and its fluctuation $C_t(k_1,k_1)$ are shown as functions of the number of condensates $K$.
Both average spectrum $S_t(k_1)$ and fluctuation $C_t(k_1,k_1)$ decay monotonically as $K$ increases, except for the fluctuation $C_t(k_1,k_1)=0$ for $K=2$.
For the double-condensate case $K=2$, the interference pattern with perfect fringe contrast, which corresponds to $S_t(k)=1/4$, is certainly observed on every snapshot with vanishing fluctuation $C_t(k_1,k_1)=0$.\cite{ref:PolkovnikovEPL-IJQI}

\subsection{Spontaneous Symmetry Breaking}
\label{sec:SSB}
The formulas for $I_t^{(2)}(k)$ and $I_t^{(4)}(k,k')$ in (\ref{eqn:I2}) and (\ref{eqn:I4}), and hence  $S_t(k)$ and $C_t(k,k')$, can be reproduced on the basis of the ``symmetry-breaking scenario'' employed in Ref.\ \refcite{ref:Hadzibabic-BECIntArray-IJQI} to analyze the experiment.
This explains well the data.
Suppose that the $\text{U}(1)$ symmetry of the system is spontaneously broken upon condensation and each condensate is described by a ``macroscopic wave function'' $\varphi_n(x)e^{i\theta_n}$ ($n=1,\ldots,K$), where $\theta_n$ is the phase of the condensate, which is randomly fixed upon condensation, varying randomly from $0$ to $2\pi$ from one realization to another.
Note that in this scenario the number of atoms in each BEC is not fixed.
The snapshot profile for a single realization of the array of condensates, with a given set of $\{\theta_1,\ldots,\theta_K\}$, is then written as
\begin{equation}
\rho(x,t)=\frac{1}{K}\left|\sum_{n=1}^K\varphi_n(x,t)e^{i\theta_n}\right|^2,
\end{equation}
and the average spectrum of many snapshots for many different realizations of the array of condensates is computed as
\begin{align}
S_t(k)
&=\int_0^{2\pi}\frac{d\theta_1}{2\pi}\cdots\int_0^{2\pi}\frac{d\theta_K}{2\pi}\,|\tilde{\rho}(k,t)|^2
\nonumber\\
&=
|I_t^{(1)}(k)|^2
+\frac{1}{K^2}\mathop{\sum_{n_1=1}^K\sum_{n_2=1}^K}_{n_1\neq n_2}
|\chi_{n_1n_2}(k,t)|^2,
\end{align}
where $I_t^{(1)}(k)$ and $\chi_{n_1n_2}(k,t)$ are defined in (\ref{eqn:chi}).
This coincides with (\ref{eqn:I2}) and (\ref{eqn:SGauss}) in the large $N$ limit.
It is also the case for $I_t^{(4)}(k,k')$ and $C_t(k,k')$.
The average over the phases induces the constraints on $n_i$ and $n_i'$ as in (\ref{eqn:I4}), and the formula (\ref{eqn:I4}) is recovered once $N+1$ and $N+2$ are approximated by $N$ in the large $N$ limit.
In this way, the description on the basis of the spontaneous symmetry breaking provides a good approximation in the large $N$ limit.\cite{ref:CastinDalibard-IJQI}

From this point of view, the decay of the strength of the average spectrum $S_t(k_n)$ in (\ref{eqn:PeakFormulaS}) and that of the fluctuation $C_t(k_m,k_n)$ in (\ref{eqn:PeakFormulaC}) as functions of $K$ are understood in the following way.
If the number of the condensates $K$ is large, it is very rare that the condensates are endowed with an appropriate set of phases $\{\theta_1,\ldots,\theta_K\}$ which allow a constructive interference among the condensates.
That is why the average spectrum $S_t(k_n)$ becomes smaller for larger $K$, as shown in Fig.\ \ref{fig:Peaks}.
At the same time, snapshot patterns with less contrast become more typical for larger $K$, and the fluctuations of the fringe spectrum $C_t(k_m,k_n)$ is accordingly reduced as $K$ is increased [while its ratio to the average spectrum $S_t(k_n)$ remains finite\cite{ref:PolkovnikovEPL-IJQI}].

\section{Summary}
We have discussed the interference of an array of independent BECs, each consisting of a fixed number of atoms, without assuming the symmetry breaking.
The statistics of the snapshot interference patterns, i.e., the average fringe spectrum expected on each snapshot and its fluctuation (covariance), have been investigated.
An analytical formula for the average spectrum valid for any time and any number of atoms has been presented,\footnote{An analytical formula for the covariance is also available but is not presented here for brevity.} and concise formulas for both average spectrum and covariance valid for large time have been derived.
The processes contributing to these quantities have been clarified, and the formulas have been shown to coincide with the ones based on the symmetry-breaking scenario if the number of the atoms is large enough, which explains well the experimental data.

An interesting aspect of the present subject is to discuss snapshots, while quantum mechanics predicts result of accumulation of many experiments.
As pointed out in Ref.\ \refcite{ref:PolkovnikovEPL-IJQI}, the average spectrum analyzed in the present paper does not supply definite information about each single snapshot, since the variance is not vanishing.
It would be a challenging future subject to explore a better strategy to study the snapshot interference in a more conclusive way.

Another interesting issue would be how it is possible to experimentally discriminate the two scenarios, i.e., measurement-induced interference and spontaneously symmetry breaking, and how they are related to each other.
In principle, the difference would become prominent when the number of atoms $N$ is small, but at the same time, the visibility of interference pattern would be low with small $N$ and experimental observation would be difficult.
It would also be a challenging future subject to seek for a strategy to clarify the difference between the two scenarios.

\section*{Acknowledgments}
This work is supported by a Special Coordination Fund for Promoting Science and Technology and the Grant-in-Aid for Young Scientists (B) (No.\ 21740294) both from the Ministry of Education, Culture, Sports, Science and Technology, Japan,
by the bilateral Italian-Japanese Projects II04C1AF4E on ``Quantum Information, Computation and Communication'' of the Italian Ministry of Education, University and Research, and
by the Joint Italian-Japanese Laboratory on ``Quantum Information and Computation'' of the Italian Ministry for Foreign Affairs.

\end{document}